\documentclass[aps,prd,superscriptaddress,twoside,twocolumn,nofootinbib,showpacs]{revtex4-1}
\usepackage{amsmath,amssymb}
\usepackage{mathtools}
\usepackage{bbold}
\usepackage[usenames,dvipsnames]{xcolor}
\usepackage{braket}
\usepackage{graphicx,bm}
\usepackage{slashed}
\usepackage{ulem}
\usepackage{booktabs}
\usepackage{tensor}

\allowdisplaybreaks

\begin{document}

\title{\boldmath Nonperturbative collisional energy loss of heavy quarks in quark-gluon plasma}

\date{\today}

\author{Nikolai Kochelev}
\email{kochelev@theor.jinr.ru}
\affiliation{Institute of Modern Physics, Chinese Academy of Sciences, Lanzhou 730000, China}
\affiliation{Bogoliubov Laboratory of Theoretical Physics, Joint Institute for Nuclear Research,
Dubna, Moscow Region 141980, Russia}

\author{Hee-Jung Lee}
\email{Correspondenting author, Email address: hjl@chungbuk.ac.kr}
\affiliation{Department of Physics Education, Chungbuk National University, Cheongju,
Chungbuk 28644, Korea}

\author{Yongseok Oh}
\email{yohphy@knu.ac.kr}
\affiliation{Department of Physics, Kyungpook National University, Daegu 41566, Korea}
\affiliation{Asia Pacific Center for Theoretical Physics, Pohang, Gyeongbuk 37673, Korea}
\affiliation{Institute for Nuclear Studies and Department of Physics,
The George Washington University, Washington, DC 20052, USA}

\author{Baiyang Zhang}
\email{zhangbaiyang@impcas.ac.cn}
\affiliation{Institute of Modern Physics, Chinese Academy of Sciences, Lanzhou 730000, China}

\author{Pengming Zhang}
\email{zhpm@impcas.ac.cn}
\affiliation{Institute of Modern Physics, Chinese Academy of Sciences, Lanzhou 730000, China}

\begin{abstract}
We suggest a new mechanism for the energy loss of fast heavy quarks in quark-gluon plasma.
This mechanism is based on pion production caused by the anomalous chromomagnetic
quark-gluon-pion interaction induced by strong topological fluctuations of the gluon fields
represented by instantons.
We found that this mechanism gives a considerable contribution to the collisional energy loss 
of a heavy quark in quark-gluon plasma, which shows a nontrivial role of nonperturbative 
phenomena in strongly interacting quark-gluon plasma.
\end{abstract}


\maketitle

The large energy loss of heavy quarks observed in high energy heavy ion collisions at Relativistic
Heavy Ion Collider (RHIC) and Large Hadron Collider (LHC) is one of exciting puzzles in the physics
of quark-gluon plasma (QGP).
In spite of the significant progress in understanding of the possible mechanisms for the energy loss of
fast heavy quarks, the complete mechanism of this effect is yet to be explored.
So far, there are many suggestions on the origins of heavy quark energy loss.
The first is the collisional energy loss that was considered in the pioneering work of
Bjorken~\cite{Bjorken82}. (See also Refs.~\cite{MT03,Mustafa04,PP08}.)
Another mechanism is the radiative energy loss due to the gluon radiation by a heavy quark induced by the
interactions with QGP partons~\cite{GW93,Zakharov96,BDMPS97,BDMS98,GLV00c,AMY02,AJMS12}.
Other suggestions include the so-called dead cone effect that comes from the suppression of gluon
emission by the quark mass at small angles~\cite{DK01b}.
The detailed discussions on the different mechanisms of the heavy quark energy suppression in heavy ion
collisions can be found, for example, in recent publications~\cite{Uphoff14,CQB15,SSS15,AAAB15} and
references therein.

Although the afore-mentioned effects must be crucial to understand the energy loss phenomena,
these ideas are based on perturbative QCD (pQCD).
This means that the system is assumed to be in the domain of pQCD for heavy quark
interactions within QGP.
However, it is now widely accepted that the QGP produced in relativistic heavy ion collisions is not
weakly interacting pQCD-like QGP but is strongly interacting QGP (sQGP) with
nonperturbative interactions between quarks and gluons~\cite{Shuryak14}.
Therefore, it is natural to speculate that nonperturbative QCD effects might have a crucial role in
unravelling the heavy quark energy loss problem.
For example, it was suggested in Ref.~\cite{Zakharov14} that thermal monopoles in QGP may lead to a
large enhancement of the parton radiative energy loss in QGP.

In this article we investigate a nonperturbative mechanism for heavy quark energy loss that is related to
the anomalous chromomagnetic quark-gluon interaction induced by instantons~\cite{Kochelev98}.
Instantons, being strong topological fluctuations of gluon fields in the QCD vacuum, play an important
role in hadron physics and give strong influence to the properties of QGP.
(For a review, see, for example, Refs.~\cite{SS98b,Diakonov02}.)
Furthermore, instantons lead to various nontrivial effective interactions such as a very specific
quark-quark interaction~\cite{tHooft76}, the chromomagnetic quark-gluon interaction~\cite{Kochelev98},
and the quark-gluon-pion interaction~\cite{BPW97,Diakonov02}.
Recently, it was demonstrated that the interaction of the last type has an important role in understaning
the cross sections of the inclusive pion production in high energy proton-proton collisions~\cite{KLZZ15}.
Furthermore, it may have a nontrivial role in unpolarized and polarized gluon distributions
of nucleons at small Bjorken $x$~\cite{KLZZ15b}.
Therefore, it is natural to expect that such nonperturbative strong interactions would have a nontrivial
role in high energy heavy ion collisions as well.
The purpose of the present work is thus to investigate the role of the chromomagnetic quark-gluon
and quark-gluon-pion interactions in the heavy quark energy loss in QGP.

In Ref.~\cite{Kochelev98}, it was shown that instantons generate a new type of chromomagnetic
quark-gluon interaction as
\begin{equation}
\mathcal{L}_I = -i  \frac{g_s\mu_a}{4M_q}\, \bar q \, \sigma^{\mu\nu} t^a q \, G^{a}_{\mu\nu},
\label{Lag1}
\end{equation}
where $t^a$ is the SU(3) Gell-Mann matrices, $\mu_a$ is the anomalous quark chromomagnetic
moment, $M_q$ is the effective  quark mass in the instanton vacuum, $g_s$ is the strong coupling
constant, and $G^{a}_{\mu\nu}$ is the gluon field strength tensor.
Within the instanton model the value of $\mu_a$ is estimated as
\begin{equation}
\mu_a = -\frac{3\pi (M_q\rho_c^{})^2}{4\alpha_s(\rho_c^{-2})},
\label{AQCM1}
\end{equation}
where $\rho_c$ is the average instanton size in QCD vacuum,
$\alpha_s(\rho_c^{-2}) = g_s^2(\rho_c)/(4\pi)$ is the strong coupling constant at the scale of $\rho_c$.
We refer the details to Refs.~\cite{Diakonov02,Kochelev09}.

\begin{figure}[t]
\centering
\includegraphics[width=0.7\columnwidth]{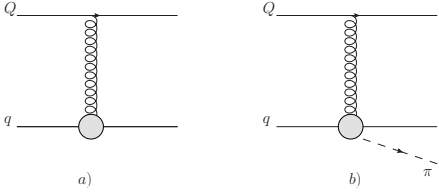}
\caption{
Diagrams for heavy quark energy loss in QGP due to (a) the quark-gluon and (b) the quark-gluon-pion
chromomagnetic interaction.
Here, $Q$ and $q$ stand for a heavy quark and a light quark, respectively.}
\label{fig:diag}
\end{figure}

However, the Lagrangian of Eq.~(\ref{Lag1}) does not respect chiral symmetry and needs
to be modified by introducing the pion field for preserving chiral symmetry~\cite{Diakonov02,BPW97}.
Then, in the single instanton approximation, the effective Lagrangian becomes
\begin{equation}
\mathcal{L}_I = -i\frac{g_s\mu_a}{4M_q} \, \bar q\sigma^{\mu\nu}t^a q \, G^{a}_{\mu\nu}
+ \frac{g_s\mu_a}{4M_qF_\pi}\bar q \sigma^{\mu\nu} t^a \gamma_5
\bm{\tau}\cdot \bm{\phi}_\pi q \, G^{a}_{\mu\nu}
\end{equation}
with the pion decay constant $F_\pi = 93~\mbox{MeV}$.
Using Eq.~(\ref{AQCM1}), it is rewritten as
\begin{eqnarray}
\mathcal{L}_I &=& i \frac{3 \pi^2 \rho_c^2 M_q}{4g_s(\rho_c)} \bar q \,\sigma^{\mu\nu} t^a q \,
G^{a}_{\mu\nu}
\nonumber\\ && \mbox{}
- \frac{3\pi^2\rho_c^2}{4g_s(\rho_c)}g_{\pi qq}^{}\,
\bar q\, \sigma^{\mu\nu} t^a \gamma_5 \bm{\tau}\cdot \bm{\phi}_\pi \, q \, G^{a}_{\mu\nu},
\label{vertex}
\end{eqnarray}
where $g_{\pi qq}^{}=M_q/F_\pi$ is the pion-quark coupling constant at zero temperature, $T=0$.
Equipped with the effective Lagrangian of the quark-gluon and quark-pion-gluon interactions,
we now consider their role in the heavy quark energy loss in QGP.
The relevant diagrams contributing to the nonperturbative heavy quark energy loss are presented in
Fig.~\ref{fig:diag}.


Our starting point is the Bjorken's formula for the collisional energy loss~\cite{Bjorken82}
with a $t$-channel exchange, which reads
\begin{equation}
\frac{dE}{dx}=\int d^3k\ n_i(k,T) \, \mathcal{F} \int d|t|\frac{d\sigma}{d|t|} \nu,
\end{equation}
where $\mathcal{F}$ is the flux factor and $\nu=E-E'$ with $E'$ being the energy of the emergent
parton.
The parton density in QGP at temperature $T$ is given by
\begin{equation}
n_i(k,T) = \frac{N_i}{(2\pi)^3} \frac{1}{\exp\left( \sqrt{k^2+m_i^2}/T \right) \pm 1},
\end{equation}
where the positive sign corresponds to the quark density ($i=q$) with $N_q = 12 \, n_f^{}$ and
$n_f^{}$ being the number of active quark flavors in QGP, while the negative sign corresponds
to the gluon density ($i=g$) with $N_g = 16$.
In our estimation, we use $n_f^{} = 2$ by considering the light $u$ and $d$ quarks.
The flux factor is $\mathcal{F} = 1-\cos\theta$, where $\theta$ is the angle between the momenta
of two incident partons.

The cross section for the diagram of Fig.~\ref{fig:diag}(a) is then calculated as
\begin{equation}
\frac{d\sigma}{dt}=\frac{\pi^3(M_q\rho_c)^2\rho_c^2F_g^2(\sqrt{|t|}\rho_c)}{8|t|},
\label{withoutpion}
\end{equation}
where $F_g(y) = 4/y^2 - 2K_2(y)$ is the instanton form factor~\cite{Kochelev09} with $K_2(y)$
being the modified Bessel function of the second kind of order~2, and we assume that, in our
nonperturbative calculation, the scale in the running strong coupling constant is determined by
the instanton size.
Therefore, the nonperturbative contribution of this diagram to the energy loss is
\begin{eqnarray}
\frac{dE^{\rm np(a)}}{dx} &=&
\frac{\pi^3(M_q\rho_c)^2\rho_c^2}{8} \int d^3k\frac{ n_i(k,T)}{2k}
\nonumber\\  && \mbox{} \times
\int_{|t|_{\rm min}}^{|t|_{\rm max}} d|t| \, F_g^2(\sqrt{|t|}\rho_c),
\end{eqnarray}
which leads to
\begin{eqnarray}
\frac{dE^{\rm np(a)}}{dx} = \frac{\pi^3(M_q\rho_c)^2\rho_c^2T^2}{16}
\int_{|t|_{\rm min}}^{|t|_{\rm max}} d|t| \, F_g^2(\sqrt{|t|}\rho_c).
\label{nonpert1}
\end{eqnarray}

For the diagram of Fig.~\ref{fig:diag}(b), consideration of the kinematics of the process gives the relation of
\begin{equation}
\nu=\frac{M_X^2-t}{2k(1-\cos\theta)},
\end{equation}
where $M_X$ is the invariant mass of the final system of the light quark and pion.
Therefore, the collisional energy loss can be rewritten as
\begin{equation}
\frac{dE}{dx} = \int \frac{d^3k}{2k} \, n_i(k,T)
\int d|t| \, \frac{d\sigma}{d|t|} \left( |t| + M_X^2 \right) .
\end{equation}

By decomposing the momentum in the longitudinal and transversal components~\cite{KLZZ15},
the differential cross section becomes
\begin{equation}
d\sigma = \frac{3g_{\pi qq}^2\rho_c^4}{2^7\pi}
\frac{F_g^2(|\bm{q}|\rho_c)}{\bm{q}^2} \, \frac{dz}{z} \, d^2\bm{q}\,
d^2 \bm{k}_\pi,
\label{cross}
\end{equation}
where $z$ is the fraction of the initial light quark momentum carried by the pion in the center of momentum
frame, $\bm{k}_\pi$ is the transverse momentum of the pion, and $\bm{q}$ is the transverse
momentum of the exchanged gluon, which leads to $t \approx - \bm{q}^2$.
Here the isospin factor $3$ is included for pion production.
By substituting $ \bm{\tilde q} = z \bm{q} - \bm{k}$ and using the relations,
\begin{equation}
\bm{\tilde q}^2 = z(1-z)M_X^2 , \quad d^2 \bm{k}_\pi = d^2 {\tilde q}=\pi z(1-z)\, dM_X^2 \, ,
\end{equation}
the integrals over $z$ and $\bm{k}_\pi$ can be performed.

Since instantons describe the subbarrier transitions between classical QCD vacua with
different topological charges, the integration over invariant mass $M_X$ of the light-quark--pion system is
restricted by the so-called sphaleron energy defined as
\begin{equation}
E_{\rm sph}=\frac{3\pi}{4\rho_c^{}\alpha_s(\rho_c^{-2})},
\end{equation}
which is the height of the potential barrier between the vacua~\cite{Diakonov02}.
The expression for the heavy quark energy loss induced by the nonperturbative quark-gluon-pion
interaction then becomes
\begin{eqnarray}
&& \frac{dE^{\rm np(b)}}{dx} =
\frac{3^3\pi^3 g_{\pi qq}^2(T)}{2^{13}} \frac{\rho_c^2 T^2}{\alpha_s^2(\rho_c^{-2})}
\nonumber \\ && \mbox{} \qquad\qquad \times
\int_{|t|_{\rm min}}^{|t|_{\rm max}} d|t|
F_g^2(\sqrt{|t|}\rho_c) \bigg(1+\frac{E_{\rm sph}^2}{2|t|}\bigg).
\label{nonpert2}
\end{eqnarray}
This result should be compared with the pQCD collisional energy loss, which is given by~\cite{PP08}
\begin{eqnarray}
\frac{dE^{\rm pert}}{dx} &=& \frac{4\pi T^2}{3} \, \alpha_s(M_D^2) \, \alpha_s(ET)
\bigg [\left(1+\frac{n_f^{}}{6} \right) \ln\frac{ET}{M_D^2}
\nonumber \\ && \mbox{} \qquad
+ \frac{2}{9}\frac{\alpha_s(M_Q^2)}{\alpha_s(M_D^2)}\ln\frac{ET}{M_Q^2}
+ c(n_f^{}) \bigg) \bigg],
\label{pert}
\end{eqnarray}
where $c(n_f^{}) \approx 0.146 \, n_f^{} + 0.05$ and $M_D$ is the Debye mass.

For our numerical estimate, we neglect the possible weak temperature-dependence of the
quark-pion coupling constant $g_{\pi qq}$ in the range of $T=(1\sim3)\,T_c$ following Ref.~\cite{EKMV93}.
This temperature range is expected to cover the temperatures achieved at the relativistic heavy
ion collisions at RHIC and LHC.
For the deconfinement temperature, we use $T_c = 150$~MeV, and the charm quark mass is
$M_c=1.3$~GeV.
The Debye mass is a function of temperature and, following the lattice QCD calculations of
Refs.~\cite{PKLSW02,DEKKL05,KZ05}, we write it as
\begin{equation}
M_D(T) \approx 3\, T.
\end{equation}
The integrals for $dE^{\rm np(a)}/dx$ and $dE^{\rm np(b)}/dx$ are performed with
$|t|_{\rm min} = M_D^2$ and $|t|_{\rm max} = ET$.

The parameters of our model are determined as follows.
For the running strong coupling constant, we use the widely used form given as~\cite{SS97c}
\begin{equation}
\alpha_s(Q^2) = \frac{1}{\beta_0} \left[ \frac{1}{\ln(Q^2/\Lambda^2)}
+ \frac{\Lambda^2}{\Lambda^2-Q^2} \right],
\label{coupling}
\end{equation}
where  $\beta_0 = (33 - 2 N_F)/12\pi$ and $\Lambda=200$~MeV for $N_F=4$.
The average size of the instanton is taken to be $\rho_c^{} = \frac13$ fm, which is supported by
both phenomenology and lattice calculations~\cite{SS98b}.
The strong coupling constant is then $\alpha_s(\rho_c^{-2}) \simeq 0.5$, which also agrees with
the value of the instanton model of Ref.~\cite{Diakonov02}.

One of the important parameter of our model is the effective quark mass in instanton vacuum.
The Lagrangians of Eqs.~(\ref{Lag1}) and (\ref{vertex}) are obtained in the effective
single instanton approximation and, in this approximation, multi-instanton effects are included
in the effective quark mass in the zero-mode-like propagator in the instanton field.
Then a careful analysis on various correlation functions within the instanton model leads to
$M_q = 86$~MeV~\cite{FS01}.
We will use this value in the present work.
But, since our results are rather dependent of the quark mass, we also discuss the results
with a higher mass, $M_q = 170$~MeV that was estimated in the earlier work of Ref.~\cite{Shuryak83}.


\begin{figure*}[t]
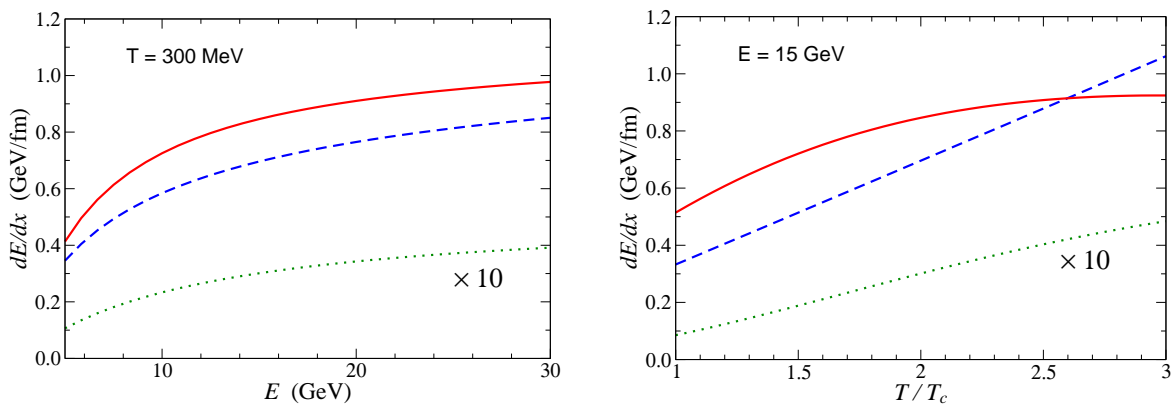

\centering
\includegraphics[width=0.85\columnwidth]{fig2a.eps}
\qquad
\includegraphics[width=0.85\columnwidth]{fig2b.eps}
\caption{(Color Online)
(a) The energy dependence of the collisional energy loss of a charm quark in QGP at $T = 2\, T_c$.
(b) The temperature dependence of the energy loss at $E=15$~GeV.
The solid lines are the nonperturbative contribution of Eq.~(\ref{nonpert1}) with the pion field,
while the dotted lines, multiplied by 10, are nonperturbative contribution of Eq.~(\ref{nonpert2})
without the pion field.
The dashed lines are the perturbative results given in Eq.~(\ref{pert}).}
\label{fig:res1}
\end{figure*}


\begin{figure*}[t]
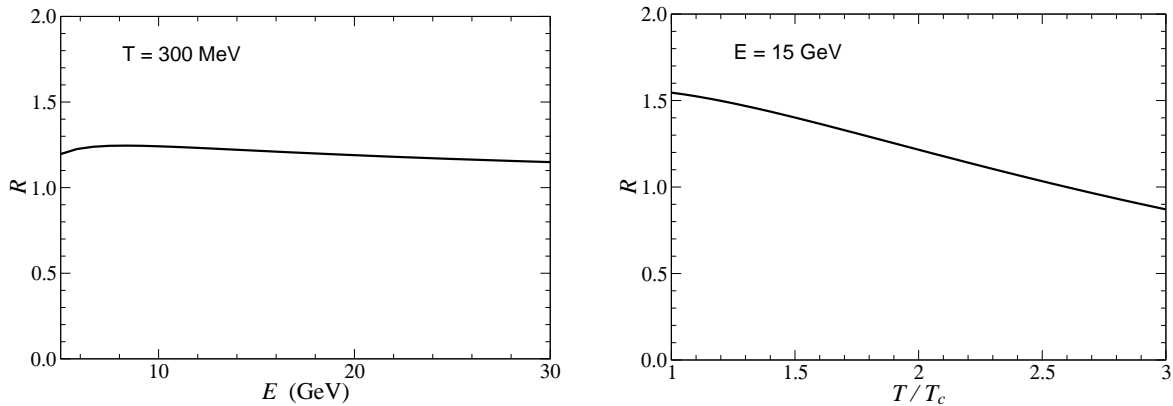

\vskip 0.5cm
\centering
\includegraphics[width=0.85\columnwidth]{fig3a.eps}
\qquad
\includegraphics[width=0.85\columnwidth]{fig3b.eps}
\caption{
(a) The energy dependence of the ratio $\mathcal{R}$ of the non-perturbative to the perturbative
contributions defined in Eq.~(\ref{eq:R}) at $T = 2\, T_c$.
(b) The temperature dependence of the ratio $\mathcal{R}$ at $E=15$~GeV.}
\label{fig:res2}
\end{figure*}

\begin{figure*}[t]
\centering
\includegraphics[width=0.9\columnwidth]{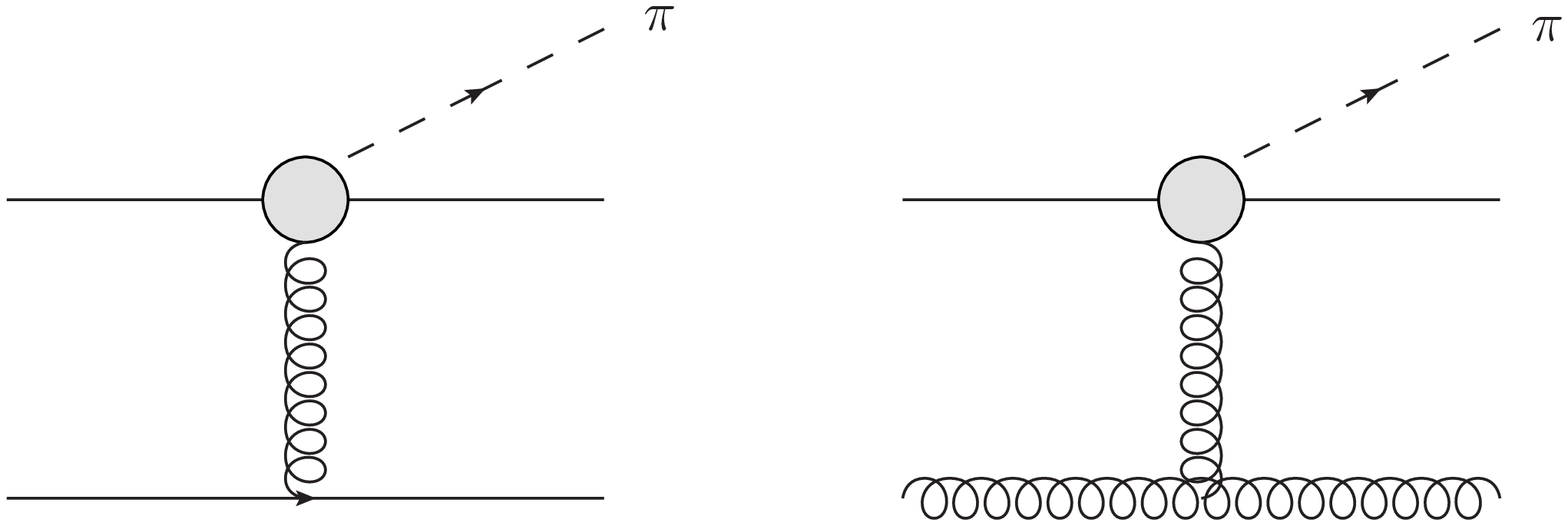}
\qquad
\includegraphics[width=0.9\columnwidth]{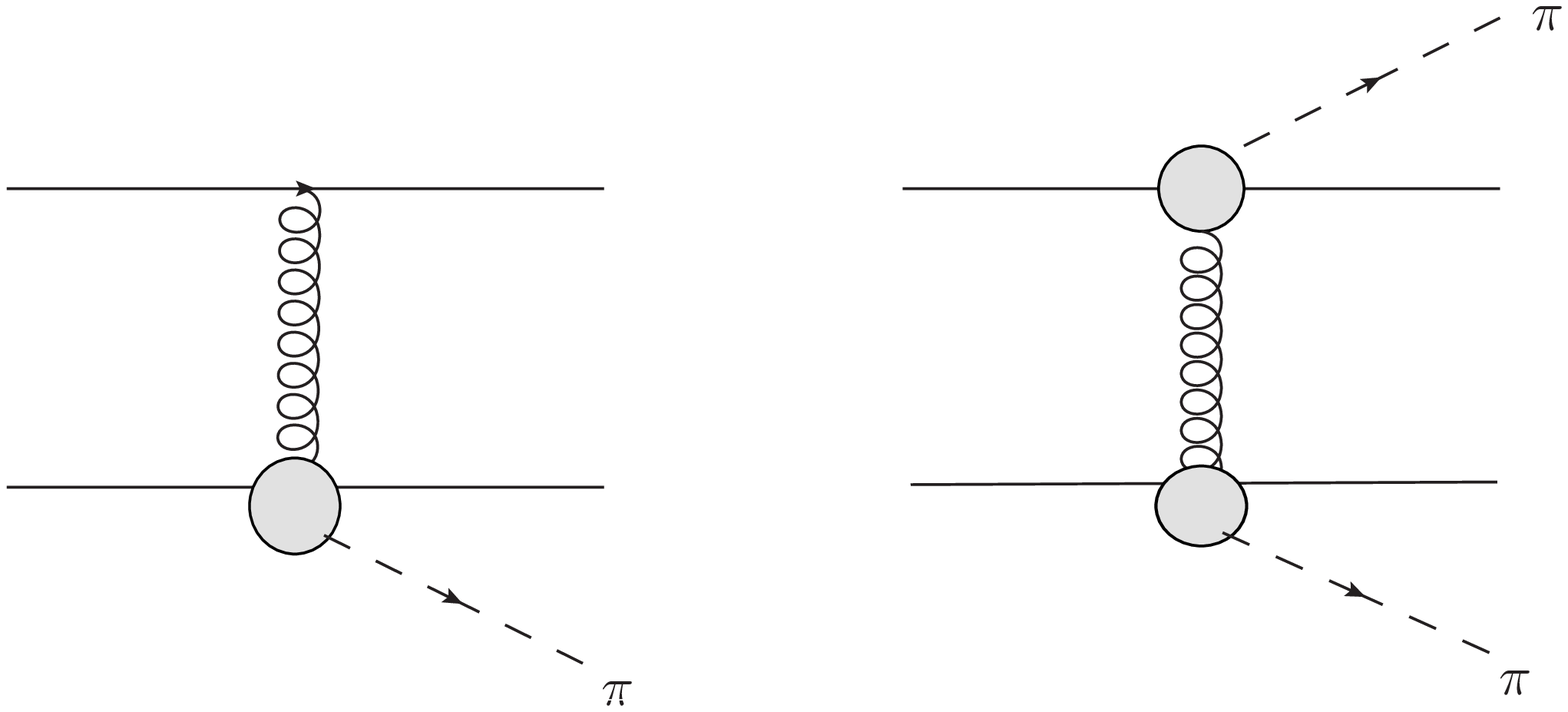}
\caption{Diagrams that can contribute to the nonperturbative light quark energy loss in QGP.}
\label{fig:diag2}
\end{figure*}

Shown in Figs.~\ref{fig:res1} and \ref{fig:res2} are the numerical results of the present calculation.
In Fig.~\ref{fig:res1}, the charm quark energy loss from the perturbative and nonperturbative parts
are shown as functions of energy and temperature.
We find that the contribution from the quark-gluon interaction without pion emission
[Fig.~\ref{fig:diag}(a)] is very small as shown by the dotted lines in Fig.~\ref{fig:res1}
and, therefore, can be safety neglected.
However, the nonperturbative  contribution with pion emission [Fig.~\ref{fig:diag}(b)] to the collisional
heavy quark energy loss is found to be similar or even larger than that of the pQCD
contribution.
To quantify the difference, in Fig.~\ref{fig:res2}, we plot the ratio $\mathcal{R}$ defined as
\begin{equation}
\mathcal{R} = \frac{dE^{\rm np(b)}}{dx} / \frac{dE^{\rm pert}}{dx}.
\label{eq:R}
\end{equation}
Figure~\ref{fig:res2} shows that $R=1.2 \sim 1.3$ in the considered region of energy and
temperature.
We also found that the energy dependence of the ratio is very weak in the energy range between
10~GeV and 30~GeV, while it has sensitive dependence on temperature.
This is because the energy dependence is determined by the cross sections that are almost saturated
in the considered energy region, while the temperature dependence is governed
by the parton density distribution.

As stated before, the radiative energy loss is expected to be one of the major sources of heavy
quark energy loss.
However, there are large uncertainties in the estimation of the radiative energy
loss~\cite{GW93,Zakharov96,BDMPS97,BDMS98,GLV00c,AMY02,AJMS12}.
In Refs.~\cite{UFXG12,MUGP12}, it was claimed that a phenomenological factor
$K =3.5$ for the ratio of the total energy loss to the pQCD collisional energy loss is needed to explain
the measured data.
This means that the contributions from the mechanisms other than pQCD contribution 
should be larger than the pQCD contribution by a factor of $2.5$.
Therefore, we conclude that a smaller radiative energy loss, namely, about 1.3 times the pQCD collisional
contribution, would be enough to resolve the heavy quark energy loss puzzle due to the 
nonperturbative contribution considered in the present work.%
\footnote{We found that the ratio $\mathcal{R}$ is rather sensitive to the effective quark mass $M_q$.
If we use $M_q = 170$~MeV, it becomes as large as $\sim 4$.
Therefore, the ambiguity in $M_q$ brings in another uncertainty originating from the
nonperturbative calculation.}

We finally make a qualitative comment on the nonperturbative energy loss of light quarks.
In this case, the virtuality of the fast quark is not large and, therefore, additional diagrams with pion
radiation from the fast light quark can give nontrivial contributions to the total energy loss.
Examples of such diagrams are shown in Fig.~\ref{fig:diag2}.
In this case, it is evident that the direct pion production from fast quarks will give enhancement of total
pion production in heavy ion collisions which, therefore, will increase the nuclear modification factor
$R_{AA}$.
However, on the other hand, the direct pion emission should lead to the energy loss of the fast quark
similar to the heavy quark case, and this will decrease $R_{AA}$.
Therefore, we have two contributions of opposite roles and the empirical $R_{AA}$ will be
determined by the competition between them.
This requires more careful and complex analyses and will be discussed elsewhere.

To summarize, we suggest a new nonperturbative mechanism for heavy quark energy loss in QGP.
It was shown that the nonperturbative chromomagnetic quark-gluon-pion interaction in
QGP may give a nontrivial contribution to the heavy quark collisional energy loss.
Therefore, this mechanism will be important to understand the mechanisms of the heavy quark energy
loss combined with other mechanisms.
Our finding again shows the important role of nonperturbative phenomena in the understanding of
the dynamics of QGP observed in high energy heavy ion collisions.

\acknowledgments

\newblock
This work was partially supported by the National Natural Science Foundation of China under Grant
Nos. 11035006 and 11175215
and by the Chinese Academy of Sciences visiting professorship for senior international scientists
under Grant No.~2013T2J0011.
Support from the MSIP of the Korean Government through the Brain Pool Program No.~121S-1-3-0318
is also gratefully acknowledged.
H.-J.L. and Y.O. were supported by the Basic Science Research Program through the National
Research Foundation of Korea (NRF) funded by the Ministry of Education under Grant Nos.
NRF-2013R1A1A2009695 and NRF-2013R1A1A2A10007294, respectively.

\end{document}